\documentclass[11pt]{article}
\usepackage{graphicx}
\usepackage{lineno}

\newcommand{\BABARPubYear}    {04}

\newcommand{\SLACPubNumber} {10657}

\def\Xpm {\ensuremath{X^\pm}\xspace} 

\RequirePackage{xspace}

\usepackage{relsize}
\def\babar{\mbox{\slshape B\kern-0.1em{\smaller A}\kern-0.1em
    B\kern-0.1em{\smaller A\kern-0.2em R}}}

\def\b     {\ensuremath{b}\xspace}

\def\piz   {\ensuremath{\pi^0}\xspace}

\def\pip   {\ensuremath{\pi^+}\xspace}
\def\pim   {\ensuremath{\pi^-}\xspace}

\def\pipm  {\ensuremath{\pi^\pm}\xspace}

\def\Kbar  {\kern 0.2em\overline{\kern -0.2em K}{}\xspace}

\def\Kz    {\ensuremath{K^0}\xspace}
\def\Kzb   {\ensuremath{\Kbar^0}\xspace}
\def\KzKzb {\ensuremath{\Kz \kern -0.16em \Kzb}\xspace}
\def\Kp    {\ensuremath{K^+}\xspace}
\def\Km    {\ensuremath{K^-}\xspace}

\def\Kmp   {\ensuremath{K^\mp}\xspace}
\def\KpKm  {\ensuremath{\Kp \kern -0.16em \Km}\xspace}
\def\KS    {\ensuremath{K^0_{\scriptscriptstyle S}}\xspace}

\def\Dbar    {\kern 0.2em\overline{\kern -0.2em D}{}\xspace}

\def\Dz      {\ensuremath{D^0}\xspace}
\def\Dzb     {\ensuremath{\Dbar^0}\xspace}
\def\DzDzb   {\ensuremath{\Dz {\kern -0.16em \Dzb}}\xspace}
\def\Dp      {\ensuremath{D^+}\xspace}
\def\Dm      {\ensuremath{D^-}\xspace}

\def\DpDm    {\ensuremath{\Dp {\kern -0.16em \Dm}}\xspace}

\def\Bbar    {\kern 0.18em\overline{\kern -0.18em B}{}\xspace}

\def\BB      {\ensuremath{B\Bbar}\xspace} 
\def\Bz      {\ensuremath{B^0}\xspace}
\def\Bzb     {\ensuremath{\Bbar^0}\xspace}
\def\BzBzb   {\ensuremath{\Bz {\kern -0.16em \Bzb}}\xspace}
\def\Bu      {\ensuremath{B^+}\xspace}
\def\Bub     {\ensuremath{B^-}\xspace}

\def\Bpm     {\ensuremath{B^\pm}\xspace}

\def\BpBm    {\ensuremath{\Bu {\kern -0.16em \Bub}}\xspace}

\def\BorBbar    {\kern 0.18em\optbar{\kern -0.18em B}{}\xspace}
\def\DorDbar    {\kern 0.18em\optbar{\kern -0.18em D}{}\xspace}
\def\KorKbar    {\kern 0.18em\optbar{\kern -0.18em K}{}\xspace}

\def\jpsi     {\ensuremath{{J\mskip -3mu/\mskip -2mu\psi\mskip 2mu}}\xspace}

\mathchardef\Upsilon="7107
\def\Y#1S{\ensuremath{\Upsilon{(#1S)}}\xspace}

\def\FourS {\Y4S}

\mathchardef\Deltares="7101
\mathchardef\Xi="7104
\mathchardef\Lambda="7103
\mathchardef\Sigma="7106
\mathchardef\Omega="710A

\def\Deltabar{\kern 0.25em\overline{\kern -0.25em \Deltares}{}\xspace}
\def\Lbar{\kern 0.2em\overline{\kern -0.2em\Lambda\kern 0.05em}\kern-0.05em{}\xspace}
\def\Sigbar{\kern 0.2em\overline{\kern -0.2em \Sigma}{}\xspace}
\def\Xibar{\kern 0.2em\overline{\kern -0.2em \Xi}{}\xspace}
\def\Obar{\kern 0.2em\overline{\kern -0.2em \Omega}{}\xspace}
\def\Nbar{\kern 0.2em\overline{\kern -0.2em N}{}\xspace}
\def\Xb{\kern 0.2em\overline{\kern -0.2em X}{}\xspace}

\def\BR         {{\ensuremath{\cal B}\xspace}}

\def\mes        {\mbox{$m_{\rm ES}$}\xspace}

\newcommand{\tev}{\ensuremath{\mathrm{\,Te\kern -0.1em V}}\xspace}
\newcommand{\gev}{\ensuremath{\mathrm{\,Ge\kern -0.1em V}}\xspace}
\newcommand{\mev}{\ensuremath{\mathrm{\,Me\kern -0.1em V}}\xspace}
\newcommand{\kev}{\ensuremath{\mathrm{\,ke\kern -0.1em V}}\xspace}
\newcommand{\ev}{\ensuremath{\mathrm{\,e\kern -0.1em V}}\xspace}
\newcommand{\gevc}{\ensuremath{{\mathrm{\,Ge\kern -0.1em V\!/}c}}\xspace}
\newcommand{\mevc}{\ensuremath{{\mathrm{\,Me\kern -0.1em V\!/}c}}\xspace}
\newcommand{\gevcc}{\ensuremath{{\mathrm{\,Ge\kern -0.1em V\!/}c^2}}\xspace}
\newcommand{\mevcc}{\ensuremath{{\mathrm{\,Me\kern -0.1em V\!/}c^2}}\xspace}

\def\mus  {\ensuremath{\rm \,\mus}\xspace}

\def\mus        {\ensuremath{\,\mu{\rm s}}\xspace}    

\def\to                 {\ensuremath{\rightarrow}\xspace}

\def\pep2{PEP-II}

\newcommand{\dedx}{\ensuremath{\mathrm{d}\hspace{-0.1em}E/\mathrm{d}x}\xspace}

\def\gsim{{~\raise.15em\hbox{$>$}\kern-.85em
          \lower.35em\hbox{$\sim$}~}\xspace}
\def\lsim{{~\raise.15em\hbox{$<$}\kern-.85em
          \lower.35em\hbox{$\sim$}~}\xspace}

\xspace

\def\jetset74   {\mbox{\tt Jetset \hspace{-0.5em}7.\hspace{-0.2em}4}\xspace}

\def\sigcut{$|m(\jpsi\pipm\piz)-3872 \mevcc|<12\mevcc$}
\def\sidcut{$48<|m(\jpsi\pipm\piz)-3872 \mevcc|<72\mevcc$}
\def\bpmmode{$B^\pm \rightarrow J/\psi \pi^\pm \pi^0 K_S^0$}
\def\b0bmode{$\bar{B}^0/B^0 \rightarrow J/\psi \pi^\pm \pi^0 K^\mp$}

\long\def\inst#1{\par\nobreak\kern 4pt\nobreak
    {\it #1}\par\vskip 10pt plus 3pt minus 3pt}

\begin{document}
{\pagestyle{empty}

\begin{flushright}
\babar-CONF-\BABARPubYear/10 \\
SLAC-PUB-\SLACPubNumber \\
July 2004 \\
\end{flushright}

\par\vskip 5cm

\begin{center}
\Large \bf 
\boldmath{ }
Search for an $\mathbf{\it{X}}$(3872) 
Charged Partner
in the Decay Mode
$\mathbf{\it{X^-} \rightarrow \it{J}/}$\boldmath{$\psi \pi^- \pi^0$}
in the $\mathbf{\it{B}}$ Meson Decays
$\mathbf{\it{B^0} \rightarrow \it{X^- K^+}}$ 
and
$\mathbf{\it{B^-} \rightarrow \it{X^- K^0_S}}$ 

\end{center}
\bigskip

\begin{center}
\large The \babar\ Collaboration\\
\mbox{ }\\
\today
\end{center}
\bigskip \bigskip

\begin{center}
\large \bf Abstract
\end{center}
We report on the search for
a charged partner of the $X(3872)$ in
the decay
$B \rightarrow X^\pm K$, 
$X^\pm \rightarrow J/\psi \pi^\pm \pi^0$,
using {213} million $\ensuremath{B\Bbar}\xspace$ events collected at the
$\ensuremath{\Upsilon{(4S)}}\xspace$ resonance with the
$\mbox{\slshape B\kern-0.1em{\smaller A}\kern-0.1em B\kern-0.1em{\smaller A\kern-0.2em R}}$
detector at the PEP-II $e^+ e^-$ asymmetric-energy storage ring.
The
resulting product branching
fraction upper limits  are
$\BR (\bar{B}^0/B^0 \rightarrow X^\pm K^\mp$, 
$X^\pm \rightarrow J/\psi \pi^\pm \pi^0)$
$< 5.8 \times 10^{-6}$
and
$\BR ( B^\pm \rightarrow X^\pm K^0_S$, 
$X^\pm \rightarrow J/\psi \pi^\pm \pi^0)$
$< 11 \times 10^{-6}$
at the $90\%$ confidence level. 
All results are preliminary.

\vfill
\begin{center}

Submitted to the 32$^{\rm nd}$ International Conference on High-Energy Physics, ICHEP 04,\\
16 August---22 August 2004, Beijing, China

\end{center}

\vspace{1.0cm}
\begin{center}
{\em Stanford Linear Accelerator Center, Stanford University, 
Stanford, CA 94309} \\ \vspace{0.1cm}\hrule\vspace{0.1cm}
Work supported in part by Department of Energy contract DE-AC03-76SF00515.
\end{center}

\newpage
} 

\begin{center}
\small

The \babar\ Collaboration,
\bigskip

%
B.~Aubert,
R.~Barate,
D.~Boutigny,
F.~Couderc,
J.-M.~Gaillard,
A.~Hicheur,
Y.~Karyotakis,
J.~P.~Lees,
V.~Tisserand,
A.~Zghiche
\inst{Laboratoire de Physique des Particules, F-74941 Annecy-le-Vieux, France }
% ...

\end{center}\newpage

\section{INTRODUCTION}
\label{sec:Introduction}

\indent \indent Since the 
discovery of the $X(3872)$ by the
Belle Collaboration~\cite{x3872-belle}, 
there have been experimental confirmations from
the CDF~\cite{x3872-cdf}, D0~\cite{x3872-d0} and 
$\babar$~\cite{x3872-babar} Collaborations.
Numerous theoretical explanations 
have been proposed for 
this high-mass, narrow-width state
decaying into $J/\psi \pi^+ \pi^-$. 
The possibilities~\cite{theory-general}
include
a charmonium state~\cite{theory-charmonium},
a meson molecular state~\cite{theory-molecule},
and 
a hybrid charmonium state~\cite{theory-hybrid}.
The Cornell potential model~\cite{eichten} predicts 
a charmonium state, previously unseen,
with quantum numbers $n^{2S+1}L_{J}$=$1^3D_2$ and $J^{PC}=2^{--}$.
Also this charmonium state should be a narrow width state with a 3.830 GeV
mass and have a large 
radiative transition rate, $X(3872)\rightarrow \gamma \chi_{c1}$, 
which has not been observed by Belle~\cite{x3872-belle}.  
Since the measured $X(3872)$ mass is very close
to the $D^{*0}\bar{D}^0$ mass threshold, another attractive
possibility is a molecular 
model which is a  bound state of mesons.
If such states exist, then bound states
of charged and neutral mesons 
or charged molecular states are plausible.

The $\pi^+ \pi^-$ mass distributions from the $X(3872)$ decay
measured by Belle~\cite{x3872-belle} 
and $\babar$~\cite{x3872-babar}, both
peak 
near the kinematic upper limit and may be
consistent with the decay of $\rho^0 \rightarrow\pi^+ \pi^-$.
However, due to limited statistics a spin-parity analysis 
has not been performed. 
If indeed, the observed decay is $X(3872)\rightarrow J/\psi \rho^0$,
then a charged partner, $X(3872)^\pm \rightarrow J/\psi \rho^\pm$, 
may exist.
Assuming the $X$ charged partner is a member of an isotriplet 
and isospin is conserved in the $B$ decays,
the decay rate of
$B\rightarrow X^\pm K$ should be twice that of
$B\rightarrow X^0 K$. 
This would 
make experimental detection of
the $X^\pm$ quite favorable.
To test this conjecture, the $\babar$ collaboration has performed
a search, presented in this paper, for the $B$-meson decays, 
$\bar{B}^0/B^0 \rightarrow X^\pm K^\mp$
and
$B^\pm \rightarrow X^\pm K^0_S$,
where
$X^\pm \rightarrow J/\psi \pi^\pm \pi^0$.

\section{THE \babar\ DETECTOR AND DATASET}
\label{sec:babar}

\indent \indent The data used in this analysis 
correspond to a total integrated luminosity of 
$193$ fb$^{-1}$ 
taken on the $\FourS$
resonance, producing a sample 
of $213.2\pm $2.3 million $\BB$ events ($N_{\BB
}$).
Data
were collected at
the PEP-II asymmetric-energy $e^{+}e^{-}$ storage ring with the $\babar$ detector, 
which is described in detail elsewhere~\cite{babar-det}. The $\babar$
detector includes a silicon vertex tracker (SVT) and a 
drift chamber (DCH) in a 1.5 Tesla solenoidal magnetic field 
to detect charged particles and measure their momenta and
energy loss (dE/dx). Photons, electrons, and neutral hadrons are detected in a
CsI(Tl)-crystal electromagnetic calorimeter (EMC). 
An internally reflecting ring-imaging
Cherenkov detector (DIRC) provides 
particle
identification 
information that is complementary to that from dE/dx.
Penetrating
muons and neutral hadrons are identified  
by resistive-plate chambers 
in the steel flux return (IFR).
Preliminary 
track-selection criteria in this analysis follow previous
$\babar$ analyses~\cite{babar-charmonium}
and a detailed explanation of particle identification 
(PID) is given elsewhere~\cite{babar-charmonium},~\cite{kaon-pid}.

\section{ANALYSIS METHOD}
\label{sec:Analysis}
\indent \indent This analysis commences with charged and neutral track selections.
Each charged track candidate is required to have at least 12 DCH 
hits and transverse momentum greater than 100 MeV/$c$. 
If it is not associated with a $K^0_{S}$ decay that track candidate
must 
originate near the nominal beam spot. 

A charged-kaon or  -pion candidate is  selected on the basis of
dE/dx 
information from the SVT and DCH and the Cherenkov angle measured by the DIRC.
An electron candidate is required to have a good match 
between the expected and measured energy loss ($\dedx$) in the DCH, and 
between the expected and measured Cherenkov angle in the DIRC. 
The ratio of EMC
shower energy to DCH momentum, and the number of 
EMC crystals associated with the track
candidate must be appropriate for an electron. 
A muon is  selected on the basis of energy deposited in 
the EMC, the number and distribution of hits in the IFR, 
the match between the IFR hits and the extrapolation of 
the DCH track into the IFR, and the depth of penetration 
of the track into the IFR.

A photon candidate is  identified from energy deposited in contiguous 
EMC crystals summed together to form a cluster which has 
total energy greater than 30 MeV and a shower shape consistent 
with that expected for an electromagnetic shower. 

The intermediate states in the 
neutral $\bar{B}^0/B^0$ $\rightarrow$ $J/\psi \pi^\pm \pi^0 K^{\mp}$,
and 
charged $B^\pm$ $\rightarrow$ $J/\psi \pi^\pm \pi^0  K_{\rm{S}}^{0}$
decay
modes used in this analysis are
$J/\psi \rightarrow e^+ e^-$, 
$J/\psi \rightarrow \mu^+  \mu^-$, 
$\pi^0 \rightarrow \gamma\gamma$, 
and $K_{S}^{0}\rightarrow \pi ^{+}\pi ^{-}$. 
They are selected to be within the mass 
intervals 
$2.95<M(e^{+}e^{-})<3.14$, 
$3.06<M\left( \mu ^{+}\mu ^{-}\right)<3.14$, 
$0.119<M(\gamma\gamma)<0.151$, 
and 
$0.4917<M\left( \pi ^{+}\pi^{-}\right)<0.5037$ $\gevcc$.
The 
$e^+e^-$ 
mass interval  
is larger than that for $\mu^+\mu^-$ 
in order
to recover
events in which part of the energy was
carried away by 
bremsstrahlung photons.
The orientation of the displacement vector
between the $K_{S}^0$ decay vertex 
and
the $J/\psi$ vertex 
in the lab 
is required 
to be 
consistent with the $K_{S}^0$ momentum direction.

The search for $B$ signal events  
utilizes two kinematic variables~\cite{babar-charmonium}: 
the energy difference $\Delta E$ between the energy of the $B$ candidate and 
the beam energy $E_{\rm{b}}^{*}$ in the $\FourS$ rest frame; and the beam-energy-substituted 
mass $\mes \equiv \sqrt{\left( E_{\rm{b}}^{*}\right)^2 -\left( p_{\rm{B}}^{*}\right) ^{2}}$, where 
$ p_{\rm{B}}^{*}$ is the reconstructed momentum of the $B$ candidate in the $\FourS$ frame.
Signal events should have 
$\mes \approx m_{\rm{B}}$, where $m_{\rm{B}}$ is the nominal mass of the $B$-meson,
and $|\Delta E|\approx 0$. 

Before the data were analyzed, the selection criteria were optimized 
and fixed
separately for the charged and neutral modes using 
a Monte Carlo (MC) simulation of signal and known backgrounds.
The number of reconstructed MC signal events ($n_{\rm{s}}^{\rm{mc}}$) and the number of
reconstructed MC background events ($n_{\rm{b}}^{\rm{mc}}$) in the signal-box were used 
to
estimate the sensitivity ratio
$n_{\rm{s}}^{\rm{mc}} / (a/2 + \sqrt{n_{\rm{b}}^{\rm{mc}}})$~\cite{punzi}, 
where $a$, the number of standard deviations of significance
desired, was set to 3.
Note the maximum of this ratio is
independent of the unknown signal
branching fraction.
This ratio was maximized by
varying the selection criteria
on  
$\Delta E$,
$\mes$, 
the $X(J/\psi \pi^\pm \pi^0)$ mass,
the $K_{S}^{0}(\pi^+ \pi^-)$ mass,
the $K_{S}^{0}$ decay length signficance,
the $\gamma\gamma$ mass,
and the
particle identification for electrons, muons
and charged kaons.
When there are more than one candidate
(on average 
there were 1.3 candidates/event) 
per event, the candidate
with the smallest absolute $\Delta E$
value was chosen.
All the following plots
are displayed with one candidate per event.

\begin{figure}[h]
\begin{center}
  \includegraphics[width=0.45\textwidth]{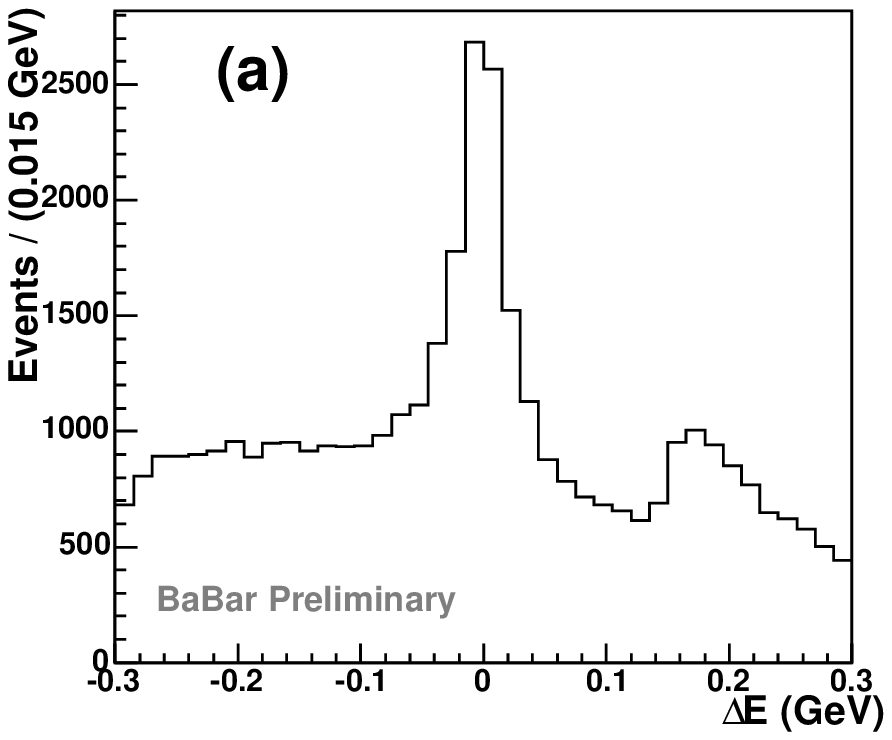}
  \includegraphics[width=0.45\textwidth]{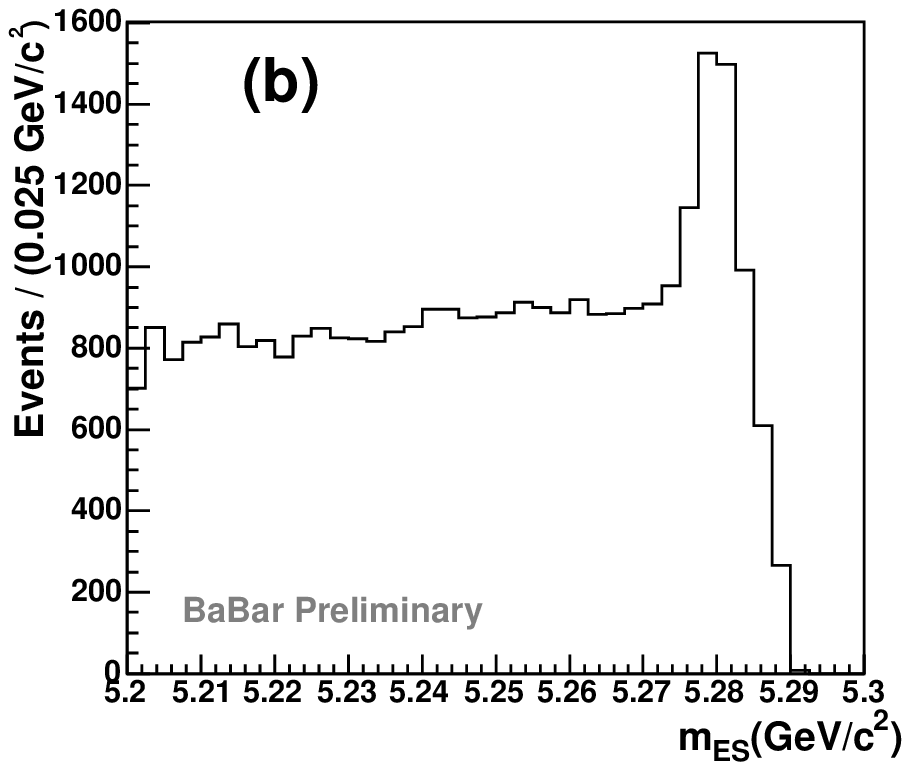}
  \caption{\label{fig:kp-mes+dele}
  The $\Delta E$ (a) and  $\mes$ (b) distributions from the
\b0bmode \ mode
after applying the optimized cuts.
  }
\end{center}
\end{figure}

\begin{figure}[h]
\begin{center}
  \includegraphics[width=0.45\textwidth]{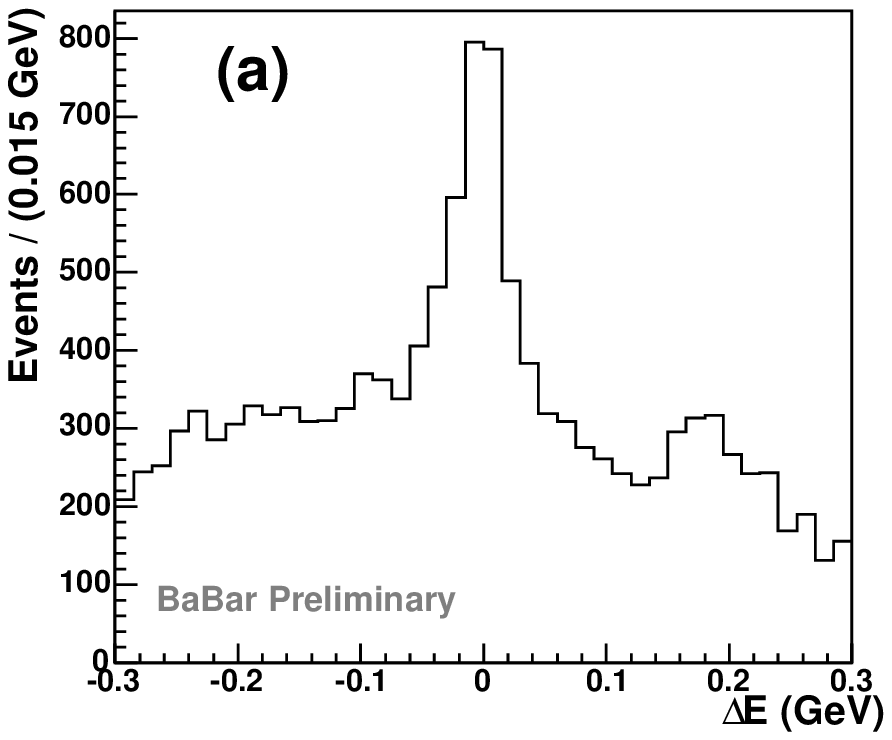}
  \includegraphics[width=0.45\textwidth]{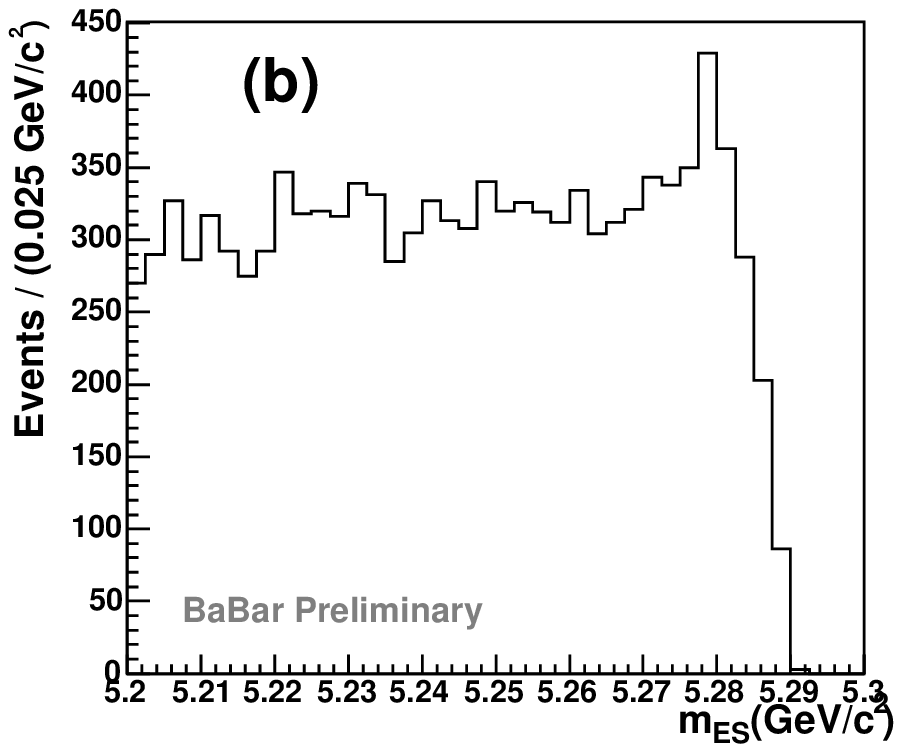}
  \caption{\label{fig:ks-mes+dele}
  The $\Delta E$ (a) and  $\mes$ (b) distributions from the
\bpmmode \ mode
after applying the optimized cuts.
  }
\end{center}
\end{figure}

\begin{figure}[h]
\begin{center}
  \includegraphics*[bb=260 250 560 460, scale=0.7]{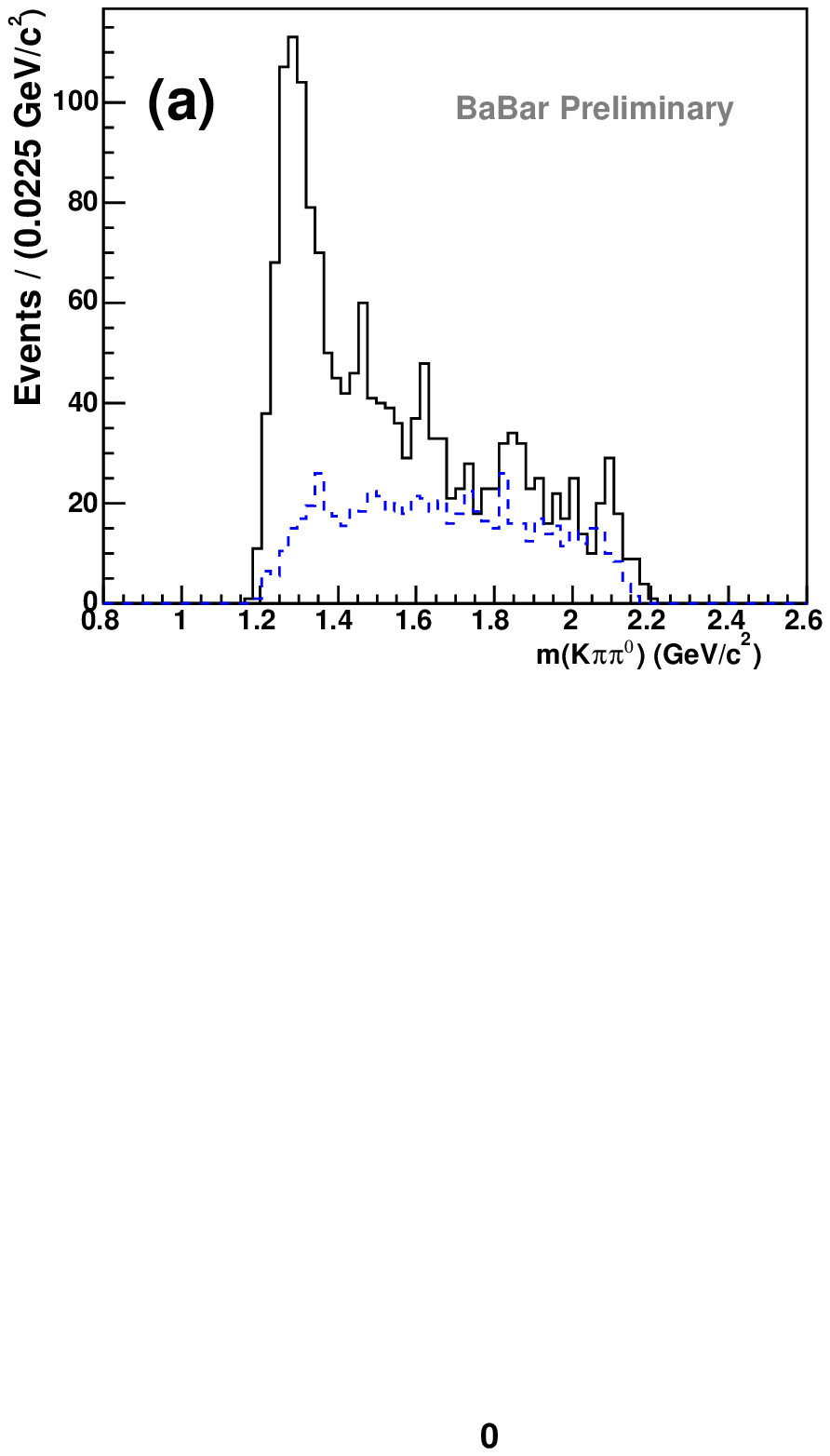}
  \includegraphics*[bb=260 250 560 460, scale=0.7]{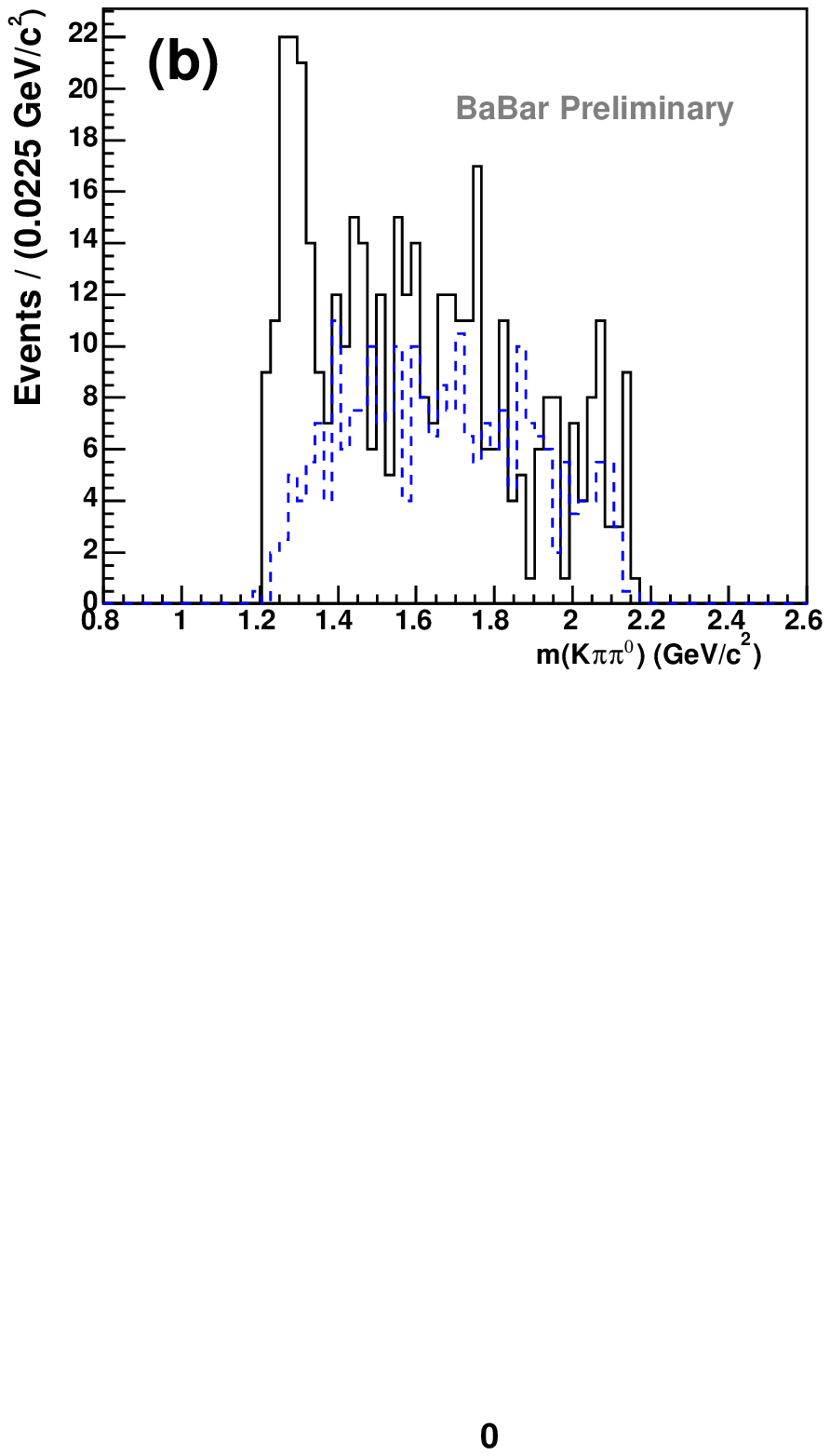}
  \caption{\label{fig:rhok-mass}
The $\pi^\pm \pi^0 K^\mp$ (a) distributions from the
\b0bmode \ mode
and
the $\pi^\pm \pi^0 K_S^0$ (b) distributions from the
\bpmmode \ mode after a signal-box and a $\rho^\pm$ mass cut.
The dashed histogram is obtained from events in a 
5.24$< \mes <$5.26 $\gevcc$ sideband.
The neutral and charged $K_1(1270)$ signals are evident
in the plots (a) and (b), respectively. 
  }
\end{center}
\end{figure}

The $\Delta E$ and $\mes$ data distributions,
after applying the optimized cuts  
for
the neutral and charged $B$ modes, are shown
in Figs.~\ref{fig:kp-mes+dele}
and 
~\ref{fig:ks-mes+dele}, respectively.
A clear signal peak is observed at zero in the $\Delta E$
distribution and near 5.279 $\gevcc$ in the $\mes$
distribution. 
The other feature in the $\Delta E$ plots is
a wide peak near 0.2 $\gevcc$ which is due
to $B \rightarrow J/\psi K^*$ events combined
with a random pion track.
The rectangular area (signal-box region) 
bounded by 
$|\mes-m_{\rm{B}}| < 5$ $\mevcc$
and $|\Delta E|<$ 20 MeV was
found to be optimal to select signal events.
Choosing events
in the signal-box region and
applying a mass cut
of $0.67<M(\pi^\pm \pi^0)<0.78$ $\gevcc$ to select
the $\rho^\pm$ mass region, the
$K^\mp \pi^\pm \pi^0$ mass distributions are
shown in Fig.~\ref{fig:rhok-mass} 
for the charged and neutral $B$ modes.
There are clear signal peaks for
$K_1^0(1270) \rightarrow K^\pm \rho^\mp$
and
$K_1^\pm(1270) \rightarrow K_S^0 \rho^\mp$
corresponding to the decays,
$B^\pm \rightarrow J/\psi K_1^\pm$
and
$B^0 \rightarrow J/\psi K_1^0$,
previously observed by Belle~\cite{belle-k1}.
In  
Fig.~\ref{fig:rhok-mass} 
the dashed histogram background estimates are
obtained using the $\mes$ 
sideband region,
$5.24<\mes<5.26$ $\gevcc$. The number of observed
$K_1$ events are consistent with the Belle measurements.

\begin{figure}
  \center{
  \includegraphics[width=0.49\textwidth]{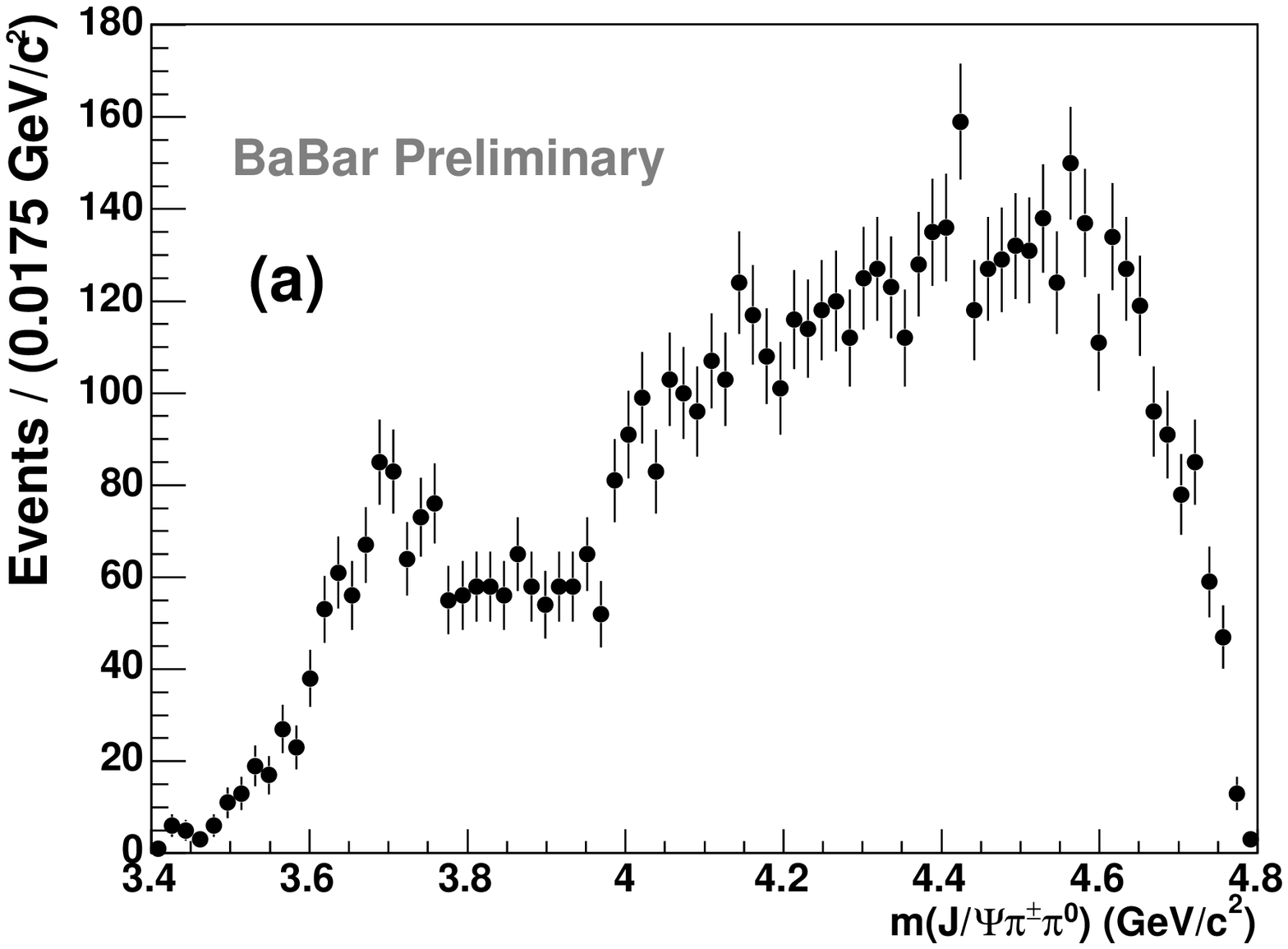}
  \includegraphics[width=0.49\textwidth]{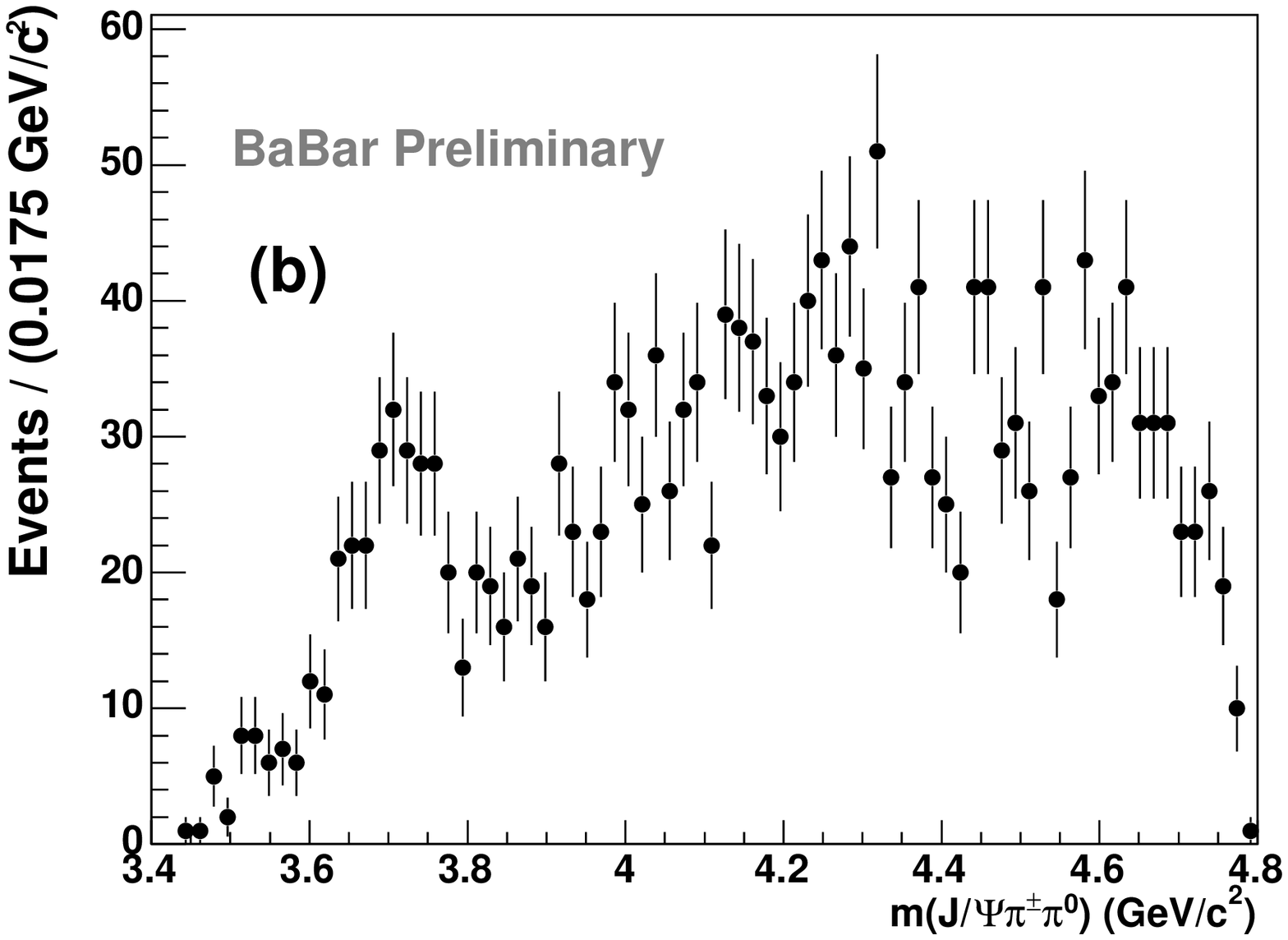}}
  \caption{\label{fig:plotX} The $\jpsi\pipm\piz$ invariant mass 
  from neutral (a), \b0bmode, and charged (b), \bpmmode, modes.}
\end{figure}

The $J/\psi \pi^\pm\pi^0$ mass spectra  
from the neutral and charged $B$ modes
are shown in Fig.~\ref{fig:plotX}.
No charged decay signal, $X^\pm \rightarrow J/\psi \pi^\pm \pi^0$,
is evident at 3.872 $\gevcc$. 
The mass spectra have backgrounds
that peak
near 3.7 $\gevcc$
and have a step near 4.0 $\gevcc$.
From MC studies we find the
peak near 3.7 $\gevcc$ is due to
$\psi(3686) \rightarrow$
$J/\psi \pi \pi$ 
decays where one pion is
exchanged with a 
random $\pi^0$. 
The
step near 4.0 $\gevcc$ 
is
caused by $B\rightarrow J/\psi K_1, K_1\rightarrow \rho K$ decays.


\section{RESULTS AND SYSTEMATIC UNCERTAINTIES}
\label{sec:upperlimits}
\indent \indent

To extract an upper limit for  $X^{\pm }\longrightarrow J/\psi \pi ^{\pm
}\pi ^{0}$, requires a search for a signal in the $J/\psi \pi ^{\pm }\pi ^{0}
$ mass, $m_{ES}$, and $\Delta E$ distributions.  A signal from $%
B\longrightarrow X^{\pm }K,$ $X^{\pm }\longrightarrow J/\psi \pi ^{\pm }\pi
^{0}$ should produce signal peaks in all three distributions. The peaking
background from non-resonant, $B\longrightarrow J/\psi \pi ^{\pm }\pi ^{0}K$%
, would produce peaks in the $m_{ES}$ and $\Delta E$ distributions and  a
flat $J/\psi \pi ^{\pm }\pi ^{0}$ mass distribution near 3.872 GeV/c$^{2}$.
The combinatoric background will not create peaks in any of the three
distributions and should produce a $m_{ES}$ distribution whose shape can be
parametrized by an ARGUS function~\cite{argus}. To estimate the number of signal events 
$(n_{S})$, we count the number of observed events $(n_{obs})$ in the signal
region and subtract the estimated number of combinatoric background events
$(n_{comb})$ and the estimated number of peaking background events $(n_{peak})$.

The number of observed events, $n_{obs}$, is obtained by counting the number
of events satisfying, $\left| m_{ES}-m_{B}\right| <5$ MeV/c$^{2}$, $\left|
\Delta E\right| <20$ MeV/c$^{2}$, and $\left| \rm{M}\left( J/\psi \pi
^{\pm }\pi ^{0}\right) -3.872\rm{ MeV/c}^{2}\right| $ $\ <$ $\ 12$ MeV/c$%
^{2}.$

\begin{figure}
  \center{\includegraphics[width=0.49\textwidth]{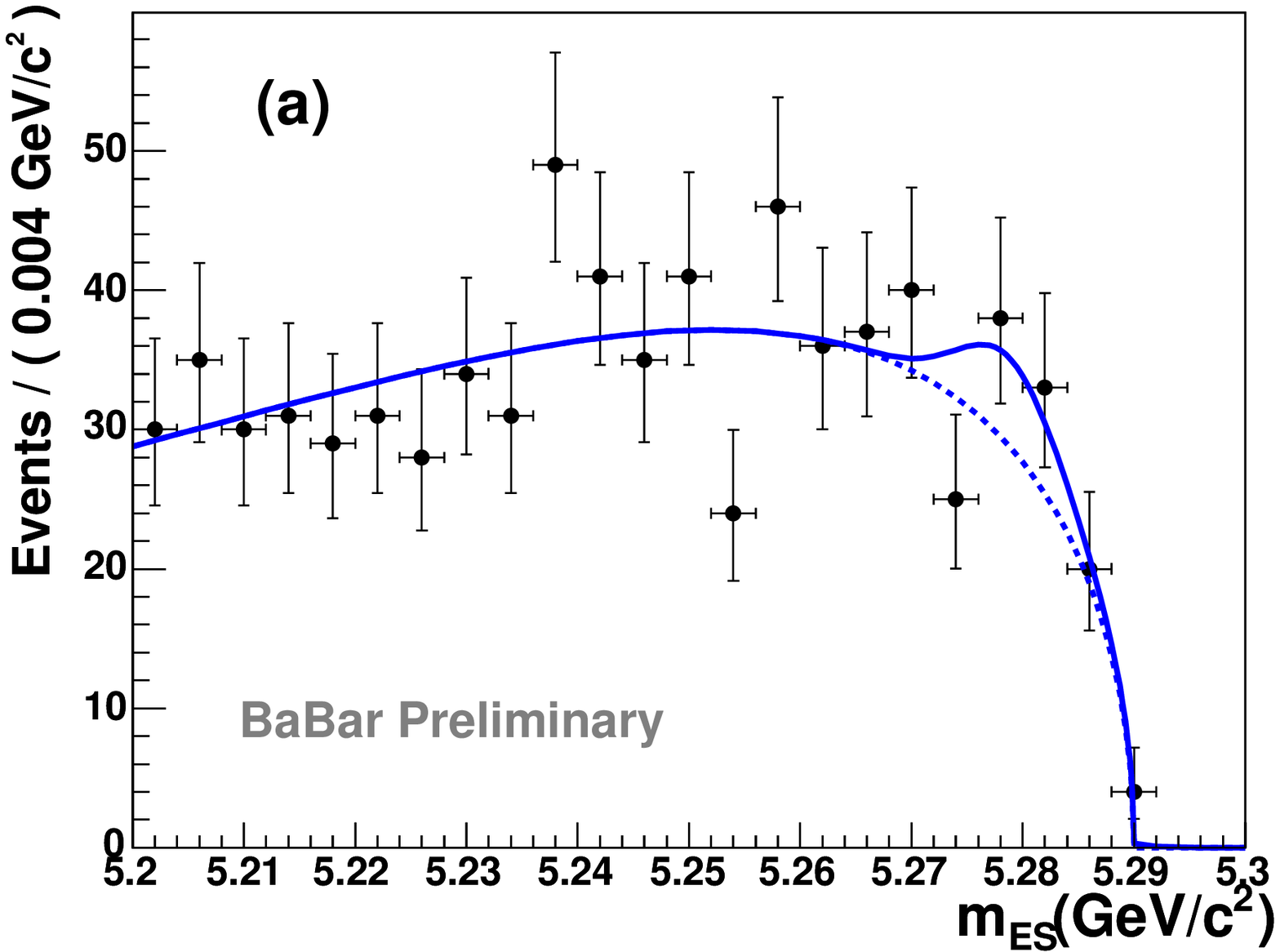}
          \includegraphics[width=0.49\textwidth]{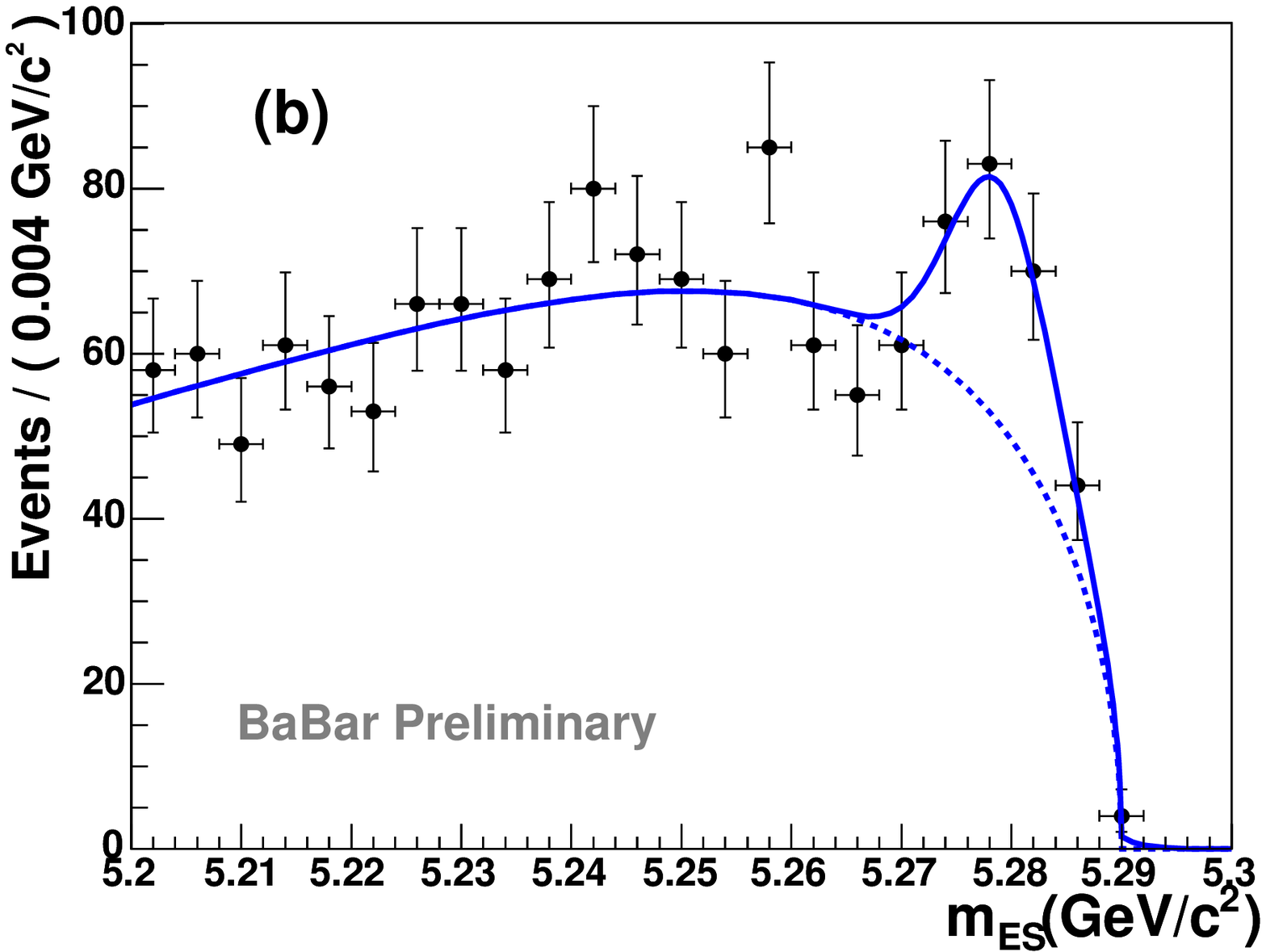}}
  \caption{\label{fig:data0-mes} 
Fitted \mes distribution for the \Bz mode with  
the $X^\pm$  signal region selection,   
\sigcut, (a)
and 
with the sideband region selection,
\sidcut, (b).
The signal region selection plot is used to estimate the
combinatoric background.
The sideband region selection plot is used to estimate the
peaking background.
The sideband region selection has twice the 
$J/\psi \pi^\pm \pi^0$ mass range
of the signal region selection. 
}
\end{figure}

\begin{figure}
  \center{\includegraphics[width=0.49\textwidth]{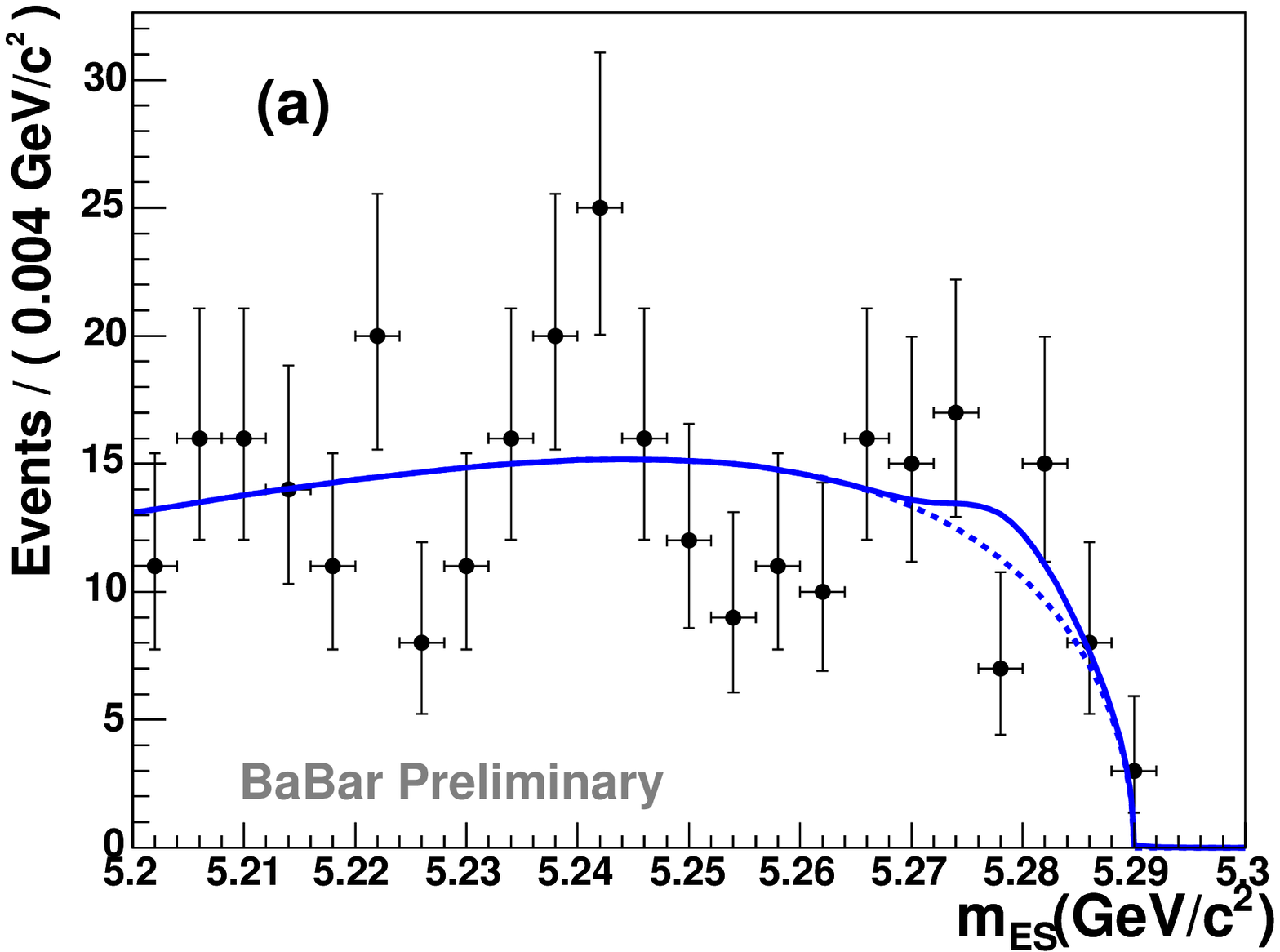}
          \includegraphics[width=0.49\textwidth]{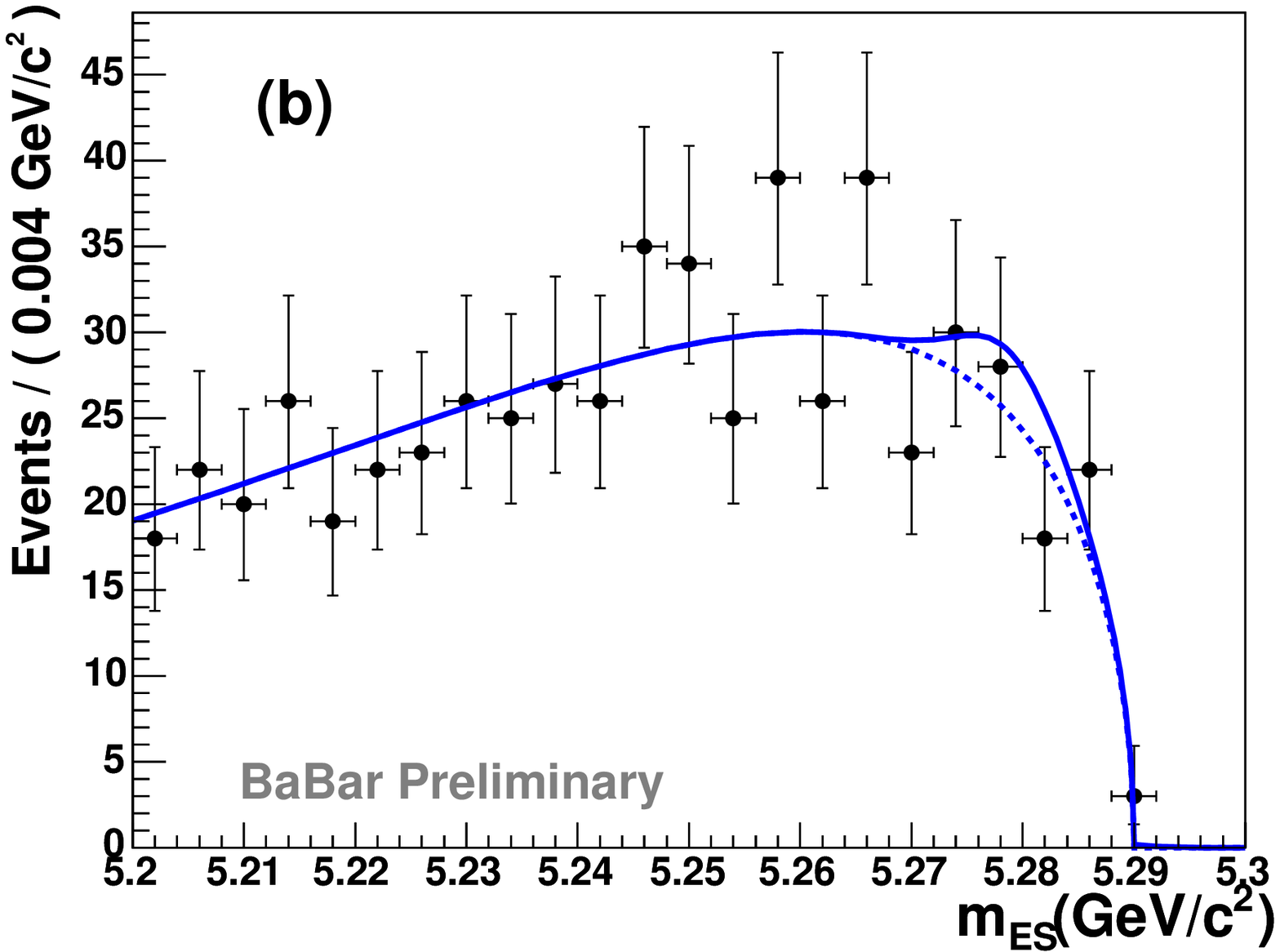}}
  \caption{\label{fig:data1-mes} 
Fitted \mes distribution for the \Bpm mode with 
the $X^\pm$  signal selection,   
\sigcut, (a)
and 
with the sideband region selection,
\sidcut, (b).
The signal region selection plot is used to estimate the
combinatoric background.
The sideband region selection plot is used to estimate the
peaking background.
The sideband region selection has twice the 
$J/\psi \pi^\pm \pi^0$ mass range
of the signal region selection. 
}
\end{figure}

The number of combinatoric background events, $n_{comb}$, is extracted from
the $m_{ES}$ distribution obtained after requiring  $\left| \Delta E\right|
<20$ MeV/c$^{2}$, and $\left| \rm{M}\left( J/\psi \pi ^{\pm }\pi
^{0}\right) -3.872\rm{ MeV/c}^{2}\right| $ $\ <$ $\ 12$ MeV/c$^{2}$. The $%
J/\psi \pi ^{\pm }\pi ^{0}$ signal band has a $24$\ MeV/c$^{2}$ wide mass 
window. 
These $m_{ES}$ distributions for the neutral and charged B modes are
separately fit with the sum of a signal Gaussian function and an ARGUS
function. The histograms with the fits are shown in 
Figs.~\ref{fig:data0-mes} and~\ref{fig:data1-mes}
for
the neutral and charged B modes, respectively. The resulting ARGUS function
is integrated over the $m_{ES}$ range,  $\left| m_{ES}-m_{B}\right| <5$ MeV/c%
$^{2}$, to produce $n_{comb}$. The error $\sigma _{comb}$ is obtained from
the fit error on the normalization of the ARGUS function. The resulting
values for $n_{comb}$ and $\sigma _{comb}$  are listed in Table 1.

The number of peaking background events, $n_{peak}$, is extracted from the $%
m_{ES}$ distribution obtained after requiring $\left| \Delta E\right| <20$
MeV/c$^{2}$, and $48<\left| \rm{M}\left( J/\psi \pi ^{\pm }\pi ^{0}\right)
-3.872\rm{ MeV/c}^{2}\right| $ $\ <$ $\ 72$ MeV/c$^{2}$. This $J/\psi \pi
^{\pm }\pi ^{0}$ sideband has a $48$\ MeV/c$^{2}$ wide mass window and is twice the
mass range of the signal band. These $m_{ES}$ distributions for the neutral
and charged B modes are separately fit with the sum of a signal Gaussian
function and an ARGUS function. The $m_{ES}$ histograms with the fits are
shown in 
Figs.~\ref{fig:data0-mes} and~\ref{fig:data1-mes}
for the neutral and charged B modes,
respectively. The estimated number of peaking background events , $n_{peak}$%
, is calculated by counting the number of events in the  $\left|
m_{ES}-m_{B}\right| <5$ MeV/c$^{2}$ $m_{ES}$ region, subtracting the number
of combinatoric events obtained from integrating the ARGUS function over the
same range, $\left| m_{ES}-m_{B}\right| <5$ MeV/c$^{2}$, and finally
dividing the result by two.  Note the Gaussian has a width that was fixed to
a value that determined from a fit to the $m_{ES}$ distribution obtained
using both the $J/\psi \pi ^{\pm }\pi ^{0}$ signal band and the $J/\psi \pi
^{\pm }\pi ^{0}$ sideband. The error $\sigma _{peak}$ is obtained by adding
in quadrature the Poisson errors on the number of events in $\left|
m_{ES}-m_{B}\right| <5$ MeV/c$^{2}$\ and the fit errors on the normalization
of the ARGUS\ function. The resulting values for $n_{peak}$ and $\sigma
_{peak}$ are listed in Table 1.

The total background ($n_b$) is the sum of the
peaking and combinatoric backgrounds and its error ($\sigma_b$)
combines in quadrature
the errors from the peaking and combinatoric backgrounds.
The backgrounds and their errors are summarized in 
Table~\ref{table-upperlimit-bf}.
\begin{table*} [!htb]
\caption{Efficiencies, number of signal-box events, 
and estimated number of background events (peaking, combinatoric, total) 
for the neutral and charged $B$ decays.
}
\begin{center}
\footnotesize{
\begin{tabular}{lccccc}
\hline\hline\\[-0.2cm]

Mode & $\epsilon$ & $n_{obs}$ & $n_{peak}\pm \sigma _{peak}$ 
& $n_{comb}\pm \sigma _{comb}$ & $n_{b}\pm \sigma _{b}$    \\ 
\hline \\[-0.2cm]

$J/\psi \pi^\pm\pi^0 K^{\mp}$ &   $10.65\% $&87 & $31.2\pm 8.0$ &$70.6\pm6.3$
&$101.8\pm 10.2$  \\ 
$J/\psi \pi^\pm\pi^0 K_{S}^{0}$ & ${8.50\%}$&31 & $ 0.6\pm 4.7$ &$27.0\pm4.0$
&$27.6\pm 6.2$ \\ 
\hline 
\end{tabular}
}
\end{center}
\label{table-upperlimit-bf}
\end{table*}

The efficiencies ($\epsilon$) for the processes,
$\bar{B}^0/B^0 \rightarrow X^\pm K^\mp$,
$X^\pm \rightarrow J/\psi \pi^\pm \pi^0$
and
$B^\pm \rightarrow X^\pm K^0_S$,
$X^\pm \rightarrow J/\psi \pi^\pm \pi^0$
are determined by MC simulation
using an $X^\pm$ signal with
zero width, mass 3.872 $\gevcc$
and 
a model
consisting of the
sequential isotropic two body decays
$B\rightarrow X^\pm K$, $X^\pm
\rightarrow J/\psi \rho^\pm$
and $\rho^\pm \rightarrow \pi^\pm \pi^0$.
Efficiencies are corrected for the small differences between 
data and MC by using well-understood control samples where results from data 
and MC are available.
These corrections
are applied to 
PID, neutral detection, and 
tracking efficiencies.
The final efficiencies for each mode are listed
in Table~\ref{table-upperlimit-bf}. 

The systematic errors include uncertainties
in the 
number of $B\bar{B}$ events in the data sample,
the secondary branching fractions,
the MC statistics,
the decay model for the generated events,
the background parametrization,
the particle identification,
the charged particle tracking,
and 
the $\pi^0$ reconstruction.
The individual uncertainties 
are given as percentages in Table~\ref{table-sys}.
The secondary branching fractions~\cite{pdg} include
 $\BR(J/\psi \rightarrow e^+e^-,\mu^+\mu^-)$=$0.1181\pm 0.0014$
and
 $\BR(K^0_S \rightarrow \pi^+\pi^-)$=$0.686\pm0.0027$.
The decay model uncertainty is estimated
by comparing the efficiencies for
phase space 
and
different decay models~\cite{pakvasa-suzuki}
with
$J^{PC}=1^{++}$ 
and
$J^{PC}=2^{--}$. 

\begin{table*} [!htb]
\caption{
Percentage Systematic Errors from the 
neutral and charged $B$ decay modes.
}
\begin{center}
\footnotesize{

\begin{tabular}{lcc}
\hline\hline
Systematic Errors(\%) & $J/\psi \pi ^{\pm }\pi ^{0}K^{\mp }$ & $J/\psi \pi
^{\pm }\pi ^{0}K_{S}^{0}$ \\ \hline
No. of $B\overline{B}$events & 1.1 & 1.1 \\ 
Branching fractions & 5.3 & 5.3 \\ 
MC statistics & 2.1 & 2.3 \\ 
MC decay model & 1.1 & 3.0 \\ 
Bkgd sideband width & 0.8 & 1.9 \\ 
Particle ID & 5.0 & 5.0 \\ 
Tracking $\pi ^{\pm }$ & 1.4 & 1.4 \\ 
Tracking $K^{\pm }$ & 1.4 & - \\ 
Tracking $K_{S}^{0}$ & - & 2.6 \\ 
Tracking $J/\psi \rightarrow e^{+}e^{-},\mu^+ \mu^-$ & 1.8 & 1.8 \\ 
$\pi ^{0}$ correction & 3.2 & 3.2 \\ \hline
TOTAL ($\sigma_{sys}$) & 8.8 & 9.7 \\ \hline
\end{tabular}

}
\end{center}
\label{table-sys}
\end{table*}

The background parametrization uncertainty is estimated by
varying the background sideband width, 
refitting the
$\mes$ distributions, and recalculating the
number of events.
The uncertainties in particle 
identification,  charged tracking
efficiency and $\pi^0$ reconstruction
efficiency are determined by studying
control samples~\cite{babar-charmonium}.
The total fractional errors ($\sigma_{sys}$) 
listed at the bottom of
Table~\ref{table-sys},
are determined by adding the 
individual contributions
in quadrature.

The probability distribution of the signal events is modeled
as a Gaussian with a mean ($n_s$) and sigma ($\sigma_s$).
For each $B$-decay mode the mean is $n_s$ = $n_{obs}$ - $n_b$ and the
sigma is $\sigma_s$ = $\sqrt{n_{obs}+\sigma_b^2}$ 
$\times \sqrt{1+\sigma_{sys}^2}$.
The systematic error is added in quadrature and scales the
errors on $n_{obs}$ and $n_b$ by the same fraction.
The results are listed in Table~\ref{table-3}.
We note the mean values, $n_s$, for the charged and
neutral modes are consistent within errors 
to zero signal events. 

The 90$\%$ confidence level (C.L.) upper limit
number of events ($N_{90}$)
is calculated using
the Gaussian probability distribution
with the assumption the number of signal events is
always greater than zero.
The integral of the distribution from zero
to $N_{90}$ will be $90\%$ of the total area above zero.
Combining 
$N_{90}$, 
$\epsilon$, 
$N_{B\bar{B}}$ events, 
and the secondary
branching fractions, 
we obtain,
\newpage
\begin{eqnarray*}
  \BR(\bar{B}^0/\Bz\to\Xpm\Kmp,\Xpm\to\jpsi\pipm\piz) &<&
           \frac{N_{90}}{\epsilon N_{\BB}\BR(\jpsi\to l^+l^-)} \\
           &=& 5.8\times10^{-6} \quad(90\%\;C.L.),\\
  \BR(\Bpm\to\Xpm\KS,\Xpm\to\jpsi\pipm\piz) &<&
\frac{N_{90}}{\epsilon N_{\BB}\BR(\jpsi\to l^+l^-)\BR(\KS\to\pip\pim)} \\
           &=& 11\times10^{-6} \quad(90\%\;C.L.).
\end{eqnarray*}
for the neutral and charged branching fraction upper limits.
For completeness we include the 
central value ($68\%$ confidence interval) 
for the
branching fraction using
the $n_s \pm \sigma_s$ values, 
\begin{eqnarray*}
  \BR(\bar{B}^0/\Bz\to\Xpm\Kmp,\Xpm\to\jpsi\pipm\piz) &=&
           \frac{n_s\pm \sigma_s}{\epsilon N_{\BB}\BR(\jpsi\to l^+l^-)} \\
           &=& (-5.5 \pm 5.2) \times10^{-6}, \\
  \BR(\Bpm\to\Xpm\KS,\Xpm\to\jpsi\pipm\piz) &=&
\frac{n_s \pm \sigma_s }{\epsilon N_{\BB}\BR(\jpsi\to l^+l^-)\BR(\KS\to\pip\pim)} \\
           &=& (2.3\pm 5.7)\times10^{-6} .
\end{eqnarray*}
The results are summarized in
Table~\ref{table-3}.
\begin{table*} [!htb]
\caption{The estimated number of signal events, 
90\% C.L. upper limit of signal events, 
the branching fraction upper limits,
and
the branching fraction $(\BR)$ for each $B$-decay.
}
\begin{center}
\footnotesize{
\begin{tabular}{lcccc}
\hline\hline\\[-0.2cm]

Mode &  $n_{s}\pm \sigma _{s}$ &$N_{90}$& $90\%$ C.L. & $\BR$    \\ 
\hline \\[-0.2cm]

$J/\psi \pi^\pm\pi^0 K^{\mp}$ &$-14.8\pm 13.9$ & $15.6$ & $< 5.8\times 10^{-6}$ & $(-5.5\pm5.2)\times 10^{-6}$\\ 
$J/\psi \pi^\pm\pi^0 K_{S}^{0}$ &$3.4\pm 8.3$&$15.9 $& $< 11\times 10 ^{-6}$ & $(2.3\pm5.7)\times 10^{-6}$ \\ 
\hline 
\end{tabular}
}
\end{center}
\label{table-3}
\end{table*}

\section{PHYSICS INTERPRETATION}
\label{sec:Physics}

\indent \indent We test the charged partner hypothesis 
at a mass of 3872 $\mevcc$
using a likelihood ratio test~\cite{pdg}.
Here we determine the ratio of the
two probabilities from the
null $(H_0)$
and 
signal $(H_1)$ 
hypotheses using our experimental 
observation of 87 events in the
signal-box.

The null hypothesis assumes the
estimated background events,
$n_b \pm \sigma_b$,
produced all the observed signal-box events.
Assuming the background probability distribution 
is a Gaussian function, we calculate a  probability
of $P(H_0)$=$7.34\times 10^{-2}$
to measure 87 or fewer events.

The isovector signal hypothesis 
predicts the product
branching fractions 
to have the ratio,
$\BR (B\rightarrow X^\pm K, X^\pm \rightarrow J/\psi \rho^\pm)$
=2
$\BR (B\rightarrow X(3872) K, X(3872) \rightarrow J/\psi \rho^0)$.
Using the $\babar$ branching fractions
$\BR (B^\pm \rightarrow X(3872) K^\pm,$ 
$X(3872) \rightarrow J/\psi \pi^+ \pi^-)$
$=(1.28\pm .41) \times 10^{-5}$
and assuming all the $\pi^+\pi^-$ decays originate
from $\rho^0$, we
expect  
$\BR (B\rightarrow X^\pm K^\mp,$ 
$X^\pm \rightarrow J/\psi \rho^\pm)$
$=(2.56\pm 0.82) \times 10^{-5}$.
This would produce 
$69\pm 23$ observed signal events
in a data sample of 213 million $B\bar{B}$ events.
The error combines the uncertainty on the branching 
fraction and the systematic error, $\sigma_{sys}$, on our
efficiency.
The probability distributions for the
signal events and the estimated background
events are modeled as two uncorrelated Gaussian functions.
The probability of observing 87 or fewer events 
with this probability distribution 
is $P(H_1)$=$1.18 \times 10 ^{-4}$.  

The likelihood ratio $(\lambda)$ test of the null hypothesis relative to
the signal hypothesis yields $\lambda$ = $P(H_0) / P(H_1)$ = $622$. 
This corresponds
to a probability of less than 1 part in 600 that the $X^\pm$ hypothesis
is correct
with the outcome of our measurement. 
Hence our result does not support the existence of charged molecular states
or charged partners of the $X(3872)$.

\section{SUMMARY}
\label{sec:Summary}
\indent \indent In conclusion, we have performed a 
search for a charged
partner of the $X(3872)$ decaying to 
$J/\psi \pi^\pm \pi^0$. Our results set 
upper limits on
the
product branching
fractions of
$\BR (\bar{B}^0/B^0 \rightarrow X^\pm K^\mp$,
$X^\pm \rightarrow J/\psi \pi^\pm \pi^0)$
$< 5.2 \times 10^{-6}$
and
$\BR ( B^\pm \rightarrow X^\pm K^0_S$,
$X^\pm \rightarrow J/\psi \pi^\pm \pi^0)$
$< 11 \times 10^{-6}$
at the $90\%$ confidence level.
We exclude the isovector $X$ hypothesis
with a likelihood ratio test and 
with our experimental results we
obtain a ratio
greater than 600  
for the null hypothesis relative to the isovector signal
hypothesis.

\section{ACKNOWLEDGMENTS}
\label{sec:Acknowledgments}
\indent \indent
We are grateful for the 
extraordinary contributions of our \pep2\ colleagues in
achieving the excellent luminosity and machine conditions
that have made this work possible.
The success of this project also relies critically on the 
expertise and dedication of the computing organizations that 
support \babar.
The collaborating institutions wish to thank 
SLAC for its support and the kind hospitality extended to them. 
This work is supported by the
US Department of Energy
and National Science Foundation, the
Natural Sciences and Engineering Research Council (Canada),
Institute of High Energy Physics (China), the
Commissariat \`a l'Energie Atomique and
Institut National de Physique Nucl\'eaire et de Physique des Particules
(France), the
Bundesministerium f\"ur Bildung und Forschung and
Deutsche Forschungsgemeinschaft
(Germany), the
Istituto Nazionale di Fisica Nucleare (Italy),
the Foundation for Fundamental Research on Matter (The Netherlands),
the Research Council of Norway, the
Ministry of Science and Technology of the Russian Federation, and the
Particle Physics and Astronomy Research Council (United Kingdom). 
Individuals have received support from 
CONACyT (Mexico),
the A. P. Sloan Foundation, 
the Research Corporation,
and the Alexander von Humboldt Foundation.

\pagebreak
\centerline{\bf \LARGE FIGURES}

\begin{figure}[h]
\begin{center}
  \includegraphics*[bb=270 0 550 245, scale=0.7]{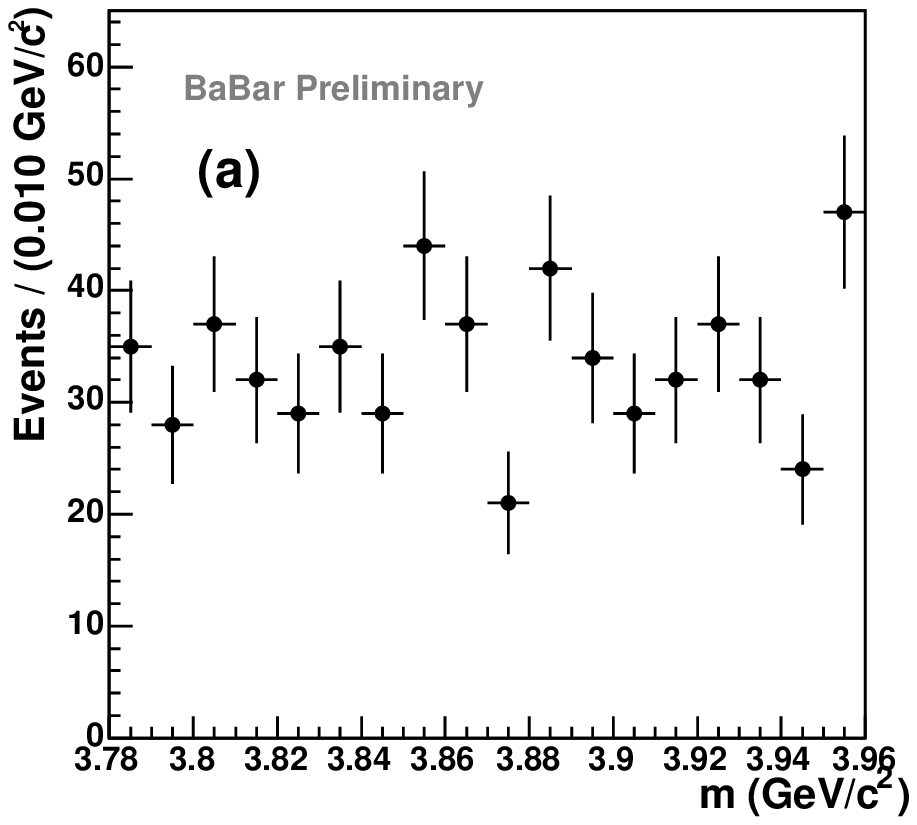}
  \includegraphics*[bb=270 0 550 245, scale=0.7]{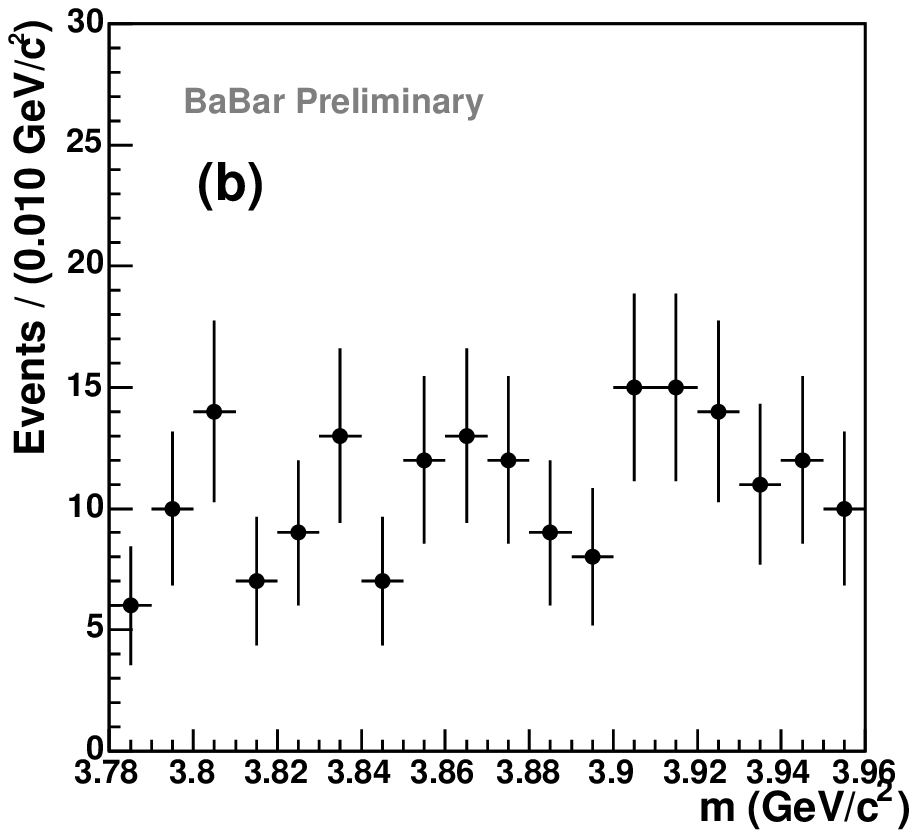}
  \caption{\label{fig:data0-m0x}The $\jpsi\pipm\piz$ invariant mass
    in 10 $\mevcc$ bins for the 
neutral (a), \b0bmode,
and 
charged (b), \bpmmode,  modes.
These plots are the same as the previous Figs. except they have finer binning.
No evidence for the charged $X(3872)$ partner is observed in either plot.
  }
\end{center}
\end{figure}

\begin{thebibliography}{99}



\bibitem{x3872-belle}
Belle Collaboration, S.K. Choi $et\ al.$, Phys. Rev. Lett. \textbf{91}, 262001 (2003).

\bibitem{x3872-cdf}
CDF Collaboration, submitted to Phys. Rev. Lett., December 5, 2003, hep-ex/0312021.


\bibitem{x3872-d0}
D0 Collaboration, submitted to Phys. Rev. Lett., May 4, 2004, hep-ex/0405004. 

\bibitem{x3872-babar}
BaBar Collaboration, submitted to Phys. Rev. Lett., June 7, 2004,  hep-ex/0406022. 

\bibitem{theory-general} 
C. Quigg, April 22, 2004, hep-ph/0403187;
E. Swanson, June 7, 2004, hep-ph/0406080;
C. Quigg, July 10, 2004, hep-ph/0407124.

\bibitem{theory-charmonium} 
E. Eichten, K. Lane, and C. Quigg, Phys. Rev. Lett. \textbf{89}, 162002 (2002);
T. Barnes and S. Godfrey, Phys.Rev.D \textbf{69}, 054008 (2004).

\bibitem{theory-molecule} 
N. Tornqvist, Phys. Lett. B \textbf{590}, 209 (2004);
M. B. Voloshin, Phys. Lett. B \textbf{579}, 316 (2004);
F. Close and P. Page, Phys. Lett. B \textbf{578}, 119 (2004);
C.Y. Wong, Phys. Rev. C \textbf{69}, 055202 (2004);
E. Braaten and M. Kusunoki, Phys. Rev. D \textbf{69}, 074004 (2004); 
E. Swanson, Phys. Lett. B \textbf{588}, 189 (2004).

\bibitem{theory-hybrid} F. Close and S. Godfrey, Phys. Lett. B \textbf{574}, 210 (2003).

\bibitem{eichten} E. Eichten, K. Lane, and C. Quigg, Phys. Rev. D  \textbf{69}, 
094019 (2004).

\bibitem{babar-det}$\babar$ Collaboration, B. Aubert $et\ al.$, Nucl. Instr. and Methods A \textbf{479}, 1 (2002).

\bibitem{babar-charmonium}$\babar$ Collaboration, B. Aubert $et\ al.$, Phys. Rev. D \textbf{65}, 032001 (2002).
This publication forms a basic reference of our analysis.
The particle identification
and tracking criteria for the photons,
electrons and muons are given in sections IIC and IID.
The photon candidates selection is described
in section VC.
The electron candidates
are required to satisfy
the ``Tight" and ``Loose" selections and
the muon candidates use
the ``Loose" selections,
as specified in Table II.
The estimate of the efficiency uncertainty in
the PID, tracking and photon detection
using control samples from data is
described in section XI.

\bibitem{kaon-pid}
$\babar$ Collaboration, B. Aubert $et\ al.$,
Phys. Rev. D \textbf{66}, 032003, (2002).
The charged kaon candidate
used a selection slightly more stringent
than that described in this reference.

\bibitem{punzi}
G. Punzi, ``Sensitivity of searches for new signals and its optimization,"
eprint physics/0308063, August 2003.

\bibitem{belle-k1}
Belle Collaboration, K. Abe $et\ al.$, Phys. Rev. Lett. \textbf{87}, 161601 (2001).

\bibitem{pdg}Particle Data Group, K. Hagiwara \textit{et al.}, Phys. Rev. D \textbf{66}, 010001 (2002),
see section 31.3.1 for a discussion of the likelihood ratio test.

\bibitem{pakvasa-suzuki}
S. Pakvasa and M. Suzuki, Phys. Lett. B \textbf{579}, 67 (2004).
Note that the authors assume the 
relative orbital angular momentum between the
$J/\psi$ and the di-pion state 
is $L=0$. This is
justified for di-pion events near the kinematic limit.  

\bibitem{argus}
The original ARGUS function is described in
H. Albrecht $et\ al.$, Phys. Lett. B \textbf{185}, 218 (1987) and
Phys. Lett. B \textbf{241}, 278 (1990).

\end{thebibliography}
\end{document}